\documentclass[onecolumn,fleqn,usenatbib]{mnras}
\usepackage{newtxtext,newtxmath}

\usepackage[T1]{fontenc}
\usepackage{lineno}
\DeclareRobustCommand{\VAN}[3]{#2}
\let\VANthebibliography\thebibliography
\def\thebibliography{\DeclareRobustCommand{\VAN}[3]{##3}\VANthebibliography}

\usepackage{graphicx}	
\usepackage{amsmath}	

\usepackage{longtable}

\newcommand{\orcid}[1]{\href{https://orcid.org/#1}{\includegraphics[width=10pt]{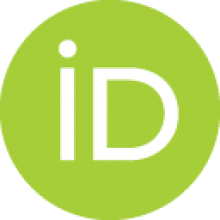}}}

\title[Comparison of Burst Properties between FRB 20190520B and FRB 20121102A]{Comparison of Burst Properties between FRB 20190520B and FRB 20121102A}

\author[F. Lyu and E.-W Liang]{
	\uppercase{Fen Lyu}\ \orcid{0000-0002-6072-3329}$^{1,2,3,4}$
 and \uppercase{En-Wei Liang}\ \orcid{0000-0002-7044-733X}$^{2}$ \thanks{E-mail: lew@gxu.edu.cn}
\\
$^{1}$Astronomical Research Center, Shanghai Science \& Technology Museum, Shanghai, 201306, China\\
$^{2}$Guangxi Key Laboratory for Relativistic Astrophysics, School of Physical Science and Technology, Guangxi University, Nanning 530004, China\\
$^{3}$Shanghai Astronomical Observatory, CAS, Nandan Road 80, Shanghai 200030, China\\
$^{4}$Key Laboratory of Modern Astronomy and Astrophysics (Nanjing University), Ministry of Education, Nanjing 210093, China
}
\date{Accepted XXX. Received YYY; in original form ZZZ}

\pubyear{2023}

\begin{document}
\label{firstpage}
\pagerange{\pageref{firstpage}--\pageref{lastpage}}
\maketitle

\begin{abstract}
A comparative analysis of the individual bursts between FRB 20190520B and FRB 20121102A is presented by compiling a sample of bursts in multiple wavelengths. It is found that the peak frequency ($\nu_p$) distribution of the bursts of FRB 20190520B illustrates four discrete peaks in $\sim1-6$ GHz and their spectral width distribution can be fitted with a log-normal function peaking at 0.35 GHz. The discrete $\nu_p$ distribution and the narrow-banded  spectral feature are analogous to FRB 20121102A. The burst duration of FRB 20190520B in the rest frame averages 10.72 ms, longer than that of FRB 20121102A by a factor 3. The specific energy ($E_{\rm \mu_{\rm c}}$) at 1.25 GHz of FRB 20190520B observed with the FAST telescope narrowly ranges in $[0.4, 1]\times 10^{38}$ erg, different from the bimodal $E_{\rm \mu_{\rm c}}$ distribution of FRB 20121102A. Assuming a Gaussian spectral profile of the bursts, our Monte Carlo simulation analysis suggests that a power-law (PL) or a cutoff power-law (CPL) energy function can comparably reproduce the $E_{\rm \mu_{\rm c}}$ distribution of FRB 20190520B. The derived energy function index of the PL model is $4.46\pm 0.17$, much steeper than that of FRB 20121102A ($1.82^{+0.10}_{-0.30}$). For the CPL model, we obtain an index of $0.47$ and a cutoff energy of $7.4\times 10^{37}$ erg. Regarding the predicted $\nu_p$ distribution in 1-2 GHz, the CPL model is more preferred than the PL model. These results indicate that FRB 20190520B and FRB 20121102A shares similar spectral properties, but their energy functions are intrinsically different.
\end{abstract}




\section{Introduction} \label{sec:intro}
Fast radio bursts (FRBs) are bright, millisecond-duration radio flashes (\citealt{2007Sci...318..777L,2019ARA/&A..57..417C,2019A&ARv..27....4P,2022A&ARv..30....2P}). Their dispersion measures (DMs) are  excess over the DM contribution of our Galaxy along the line of sight, suggesting that they are of extragalactic origin \citep{2013Sci...341...53T}. This is confirmed with the host galaxies localization of some FRBs (e.g.,\citealt{2017Natur.541...58C,2019Sci...365..565B,2019Natur.572..352R}) \footnote{http://frbhosts.org/}. The FRB phenomenon is still a great mystery, although various progenitors and radiation physics models have been proposed (see \citealt{2019PhR...821....1P} for a review). The most popular progenitor models involve compact objects, such as the magnetar \citep{2014MNRAS.442L...9L,2017ApJ...843L..26B,2020ApJ...894L..22W}, the pulsar \citep{2016MNRAS.458L..19C,2016MNRAS.462..941L,2021FrPhy..1624503L}, the magnetized neutron star \citep{2016ApJ...829...27D,2020ApJ...897L..40D}, and the strange star \citep{2021Innov...200152G}. Among over 800 FRBs reported so far \footnote{https://www.herta-experiment.org/frbstats/catalogue}, most are one-off events (non-repeating FRBs or once-off FRBs) even after long-term follow-up observations, and a small subset of FRBs sources ($\sim$ 30) exhibit repetitive behaviors (repeating FRBs, e.g., \citealt{2016Natur.531..202S,2019ApJ...885L..24C,2020ApJ...891L...6F}). The repeating FRBs tend to have longer durations and narrower bandwidths compared with the non-repeating FRBs from the observations of the first CHIME/FRB catalog \citep{2019ApJ...885L..24C,2021ApJS..257...59C,2021ApJ...923....1P}. The repetitions eliminated models concerning catastrophic events for the repeating FRBs, although they could still be potential progenitor models of non-repeating FRBs (e.g., \citealt{2013ApJ...776L..39K,2016ApJ...822L...7W}). 

FRB 20121102A and FRB 20190520B are two active repeating FRBs \citep{2014ApJ...790..101S,2016Natur.531..202S,2022Natur.606..873N}. They share similar observational features in many aspects. During the monitoring campaigns with the Five-hundred-meter Aperture Spherical Radio Telescope (FAST), they show high burst rates, i.e. $\sim 27.8\ \mathrm{ hr}^{-1}$ for FRB 20121102A \citep{2021Natur.598..267L} and $4.5_{-1.5}^{+1.9}\ \mathrm{ hr}^{-1}$ for FRB 20190520B \citep{2022Natur.606..873N}. Their host galaxies are identified as dwarf galaxies at a redshift of 0.193 for FRB 20121102A, and 0.241 for FRB 20190520B observed with the Karl G. Jansky Very Large Array (VLA; \citealp{2017Natur.541...58C,2017ApJ...834L...7T,2022Natur.606..873N}). Their rotation measure ($\mathrm{RM})$ is $\sim 10^{5}\, \mathrm{rad} \,\mathrm{m}^{-2}$, indicating that they located in an extreme and dynamic magneto-ionic circumstance (\citealt{2018Natur.553..182M,2022NatAs...6..393N}). They also show obvious depolarization \citep{2022Sci...375.1266F}. They have high DM measurements, e.g. $565.8 \pm 0.9 \mathrm{pc}\,\mathrm{cm}^{-3}$ for FRB 20121102A observed from MJD 58,724 to MJD 58,776 \citep{2021Natur.598..267L} and $1,204.7 \pm 4.0 \,\mathrm{pc} \,\mathrm{cm}^{-3}$ for the mean DM of FRB 20190520B \citep{2022Natur.606..873N}. Compact persistent radio sources (PRSs) are identified to be associated with these two FRBs \citep{2017Natur.541...58C,2022Natur.606..873N}.  

The radiation physics of FRBs is an open question. Their high brightness temperatures ($T_{\rm B} \sim 10^{35}$ K) indicate that the emission should be related to coherent mechanisms (see \citealt{2020Natur.587...45Z} for a review), such as the synchrotron maser mechanisms in relativistic shocks   \citep{2014MNRAS.442L...9L,2017ApJ...843L..26B,2020ApJ...896..142B,2019MNRAS.485.4091M,2020ApJ...899L..27M} and or the pulsar magnetosphere \citep{2020MNRAS.494.2385K,2020ApJ...899..109W}. 
It seems ubiquitous for the repeating FRBs that sub-bursts exhibit downward-drifting (e.g., \citealt{2018ApJ...863....2G,2019ApJ...885L..24C,2019ApJ...876L..23H,2020ApJ...891L...6F}). The burst spectra of the repeating FRBs are highly variable. Some bursts from repeating FRBs can be described by a Gaussian envelope (eg., \citealt{2017ApJ...850...76L,2021ApJ...920L..18A,2021ApJ...923....1P}), and the burst peak frequency differs significantly within the instrumental bandwidth
\citep{2016Natur.531..202S,2016ApJ...833..177S,2017ApJ...850...76L,2018ApJ...863....2G,2018ApJ...866..149Z,2022MNRAS.515.3577H}. It is interesting that \cite{2022ApJ...941..127L} found a signature of a discrete fringe pattern in the burst peak frequency ($\nu_{p}$) distribution of FRB 20121102A with the $C$-band (4-8 GHz) data observed by the Green Bank Telescope (GBT; \citealp{2018ApJ...863....2G,2018ApJ...866..149Z}). By extrapolating this fringe pattern to the $L$-band, they showed that the observed bimodal distribution of the observed specific burst energy ($E_{\mu \rm c}$) of FRB 20121102A at the central frequency of the FAST band \citep{2021Natur.598..267L} and the narrow spectral width distribution can be reproduced via Monte Carlo simulations, assuming that the intrinsic burst energy function is a single power law function \citep{2022ApJ...941..127L}.

We investigate whether FRB 20190520B shares a similar spectral and intrinsic burst energy function as FRB 20121102A in this paper. The paper is structured as follows. We compare the observational properties of FRB 20190520B with that of FRB 20121102A in Section 2 and then present our Monte Carlo simulation results of FRB 20190520B in Section 3. Conclusions and a discussion of our results are given in Section 4. The latest Planck flat $\Lambda$CDM cosmological model with the parameters $H_{0}$=67.7 $\mathrm{~km}$ $\mathrm{~s}^{-1}$ $\mathrm{Mpc}^{-1}$, $\Omega_{m}=0.31$ \citep{2016A/&A...594A..13P} is adopted throughout the paper.
 
\section{Comparison of the burst properties between FRB 20190520B and FRB 20121102A} \label{sec:observation}
\subsection{Sample and Data}
FRB 20190520B has been observed over a wide frequency range by different telescopes. We compile a sample of 124 bursts from literature, including 79 bursts detected with the FAST telescope in 1.05-1.45 GHz \citep{2022Natur.606..873N}, 28 bursts observed with the GBT telescope (9 bursts in 1.1-1.9 GHz and 16 bursts in 4-8 GHz, \citealp{2022arXiv220211112A}; 3 bursts in 4-8 GHz, \citealp{2022Sci...375.1266F}), 9 bursts observed with VLA (8 bursts detected in 1-4 GHz and 1 burst in 5-7 GHz; \citealp{2022Natur.606..873N}), and 8 bursts observed with the Parkes telescope in 0.704-4.032 GHz \citep{2022arXiv220308151D}.  We collect the spectral information of those bursts. Since the emission edges of a lot of bursts in the FAST sample extend beyond the observed bandpass of the FAST telescope \citep{2021ApJ...920L..18A,2022Natur.606..873N}, we cannot obtain the emission peak frequency ($\nu_p$) and spectral bandwidth ($\Delta \nu$) of these bursts. Among the FRB 20190520B bursts emitted that have been reported, the $\nu_{p}$ values are available for 11 bursts (3 GBT bursts from \citep{2022Sci...375.1266F} and 8 Parkes bursts from \citealt{2022arXiv220308151D}). The emission frequency ranges (the high-frequency edge $\mathrm{f}_{\text {high }}$ and the low-frequency edge $\mathrm{f}_{\text {low}}$) are available for 34 bursts (25 GBT bursts from \citealt{2022arXiv220211112A} and 9 VLA bursts from \citealt{2022Natur.606..873N}). We estimate the $\nu_{p}$ and $\Delta \nu$ values of these bursts with $\nu_{p}$= ($\mathrm{f}_{\text {high }}$+$\mathrm{f}_{\text {low }}$)/2 and $\Delta \nu$ = $\mathrm{f}_{\text {high }}$-$\mathrm{f}_{\text {low }}$, respectively. We finally have a sample of 45 (34) bursts that have $\nu_{p}$ ($\Delta\nu$) measurement. The temporal and spectral properties of these 124 bursts are summarized in Table \ref{collect}. 

Thousands of bursts of FRB 20121102A have been accumulated from extensive monitoring with various radio telescopes in the frequency range from 400 MHz to 8 GHz (e.g., \citealt{2016Natur.531..202S,2019ApJ...882L..18J,2021Natur.598..267L,2018ApJ...863....2G,2018ApJ...866..149Z}). We utilize the data of FRB 20121102A observed with the FAST, Arecibo in the $L$-band, and GBT telescopes in the $C$-band for our comparative analysis. The data details and their references are available in \cite{2022ApJ...941..127L}.

\subsection{Comparisons of the Burst Energy and Duration observed with the FAST telescope}
 The observed energy and burst duration are sensitive to the bandpass, the instrumental central frequency, and the detection threshold of telescopes. To avoid the instrumental selection effects, we compare the burst energy and duration observed with the FAST telescope in the $L$-band. Both FRB 20190520B and FRB 20121102A were extensively observed with the FAST telescope. We adopt the largest samples observed with the FAST telescope in a given observational campaign for our comparison, i.e. a sample of 79 bursts for FRB 20190520B \citep{2022Natur.606..873N} and a sample of 1652 bursts for FRB 20121102A \citep{2021Natur.598..267L}. The burst duration in the rest frame is calculated by $W=\omega /(1+z)$, where $\omega$ is the burst observed duration. The specific burst energy $\mathbf{E_{\rm \mu_{\rm c}}}$ at the FAST central frequency of $\mu_{\rm c}=1.25 \,\mathrm{GHz}$ is calculated with   
\begin{equation}
\mathbf{E_{\rm \mu_{\rm c}}}
\simeq
\left(10^{39} {\rm erg}\right)
\frac{4 \pi}{1+z}\left(\frac{D_{\rm L}}{10^{28} {\rm ~cm}}\right)^{2}
\left(\frac{F_{\nu}}{{\rm Jy} \cdot {\rm ms}}\right)\left(\frac{\mathbf{\mu_{\rm c}}}{\rm GHz}\right), \label{Energy}
\end{equation}
where $D_{\mathrm{L}}$ is the luminosity distance  (
$d_{\mathrm{L}}=1246.28\,\mathrm{Mpc}$ for FRB 20190520B, \citealt{2022Natur.606..873N}) and $F_{\nu}$ is the specific fluence given by $F_\nu=S_{\nu, p} \omega_{\mathrm{obs}}$, in which $S_{\nu, p}$ is the specific peak flux.
  
\begin{figure}
\centering
\includegraphics[width=0.45\linewidth]{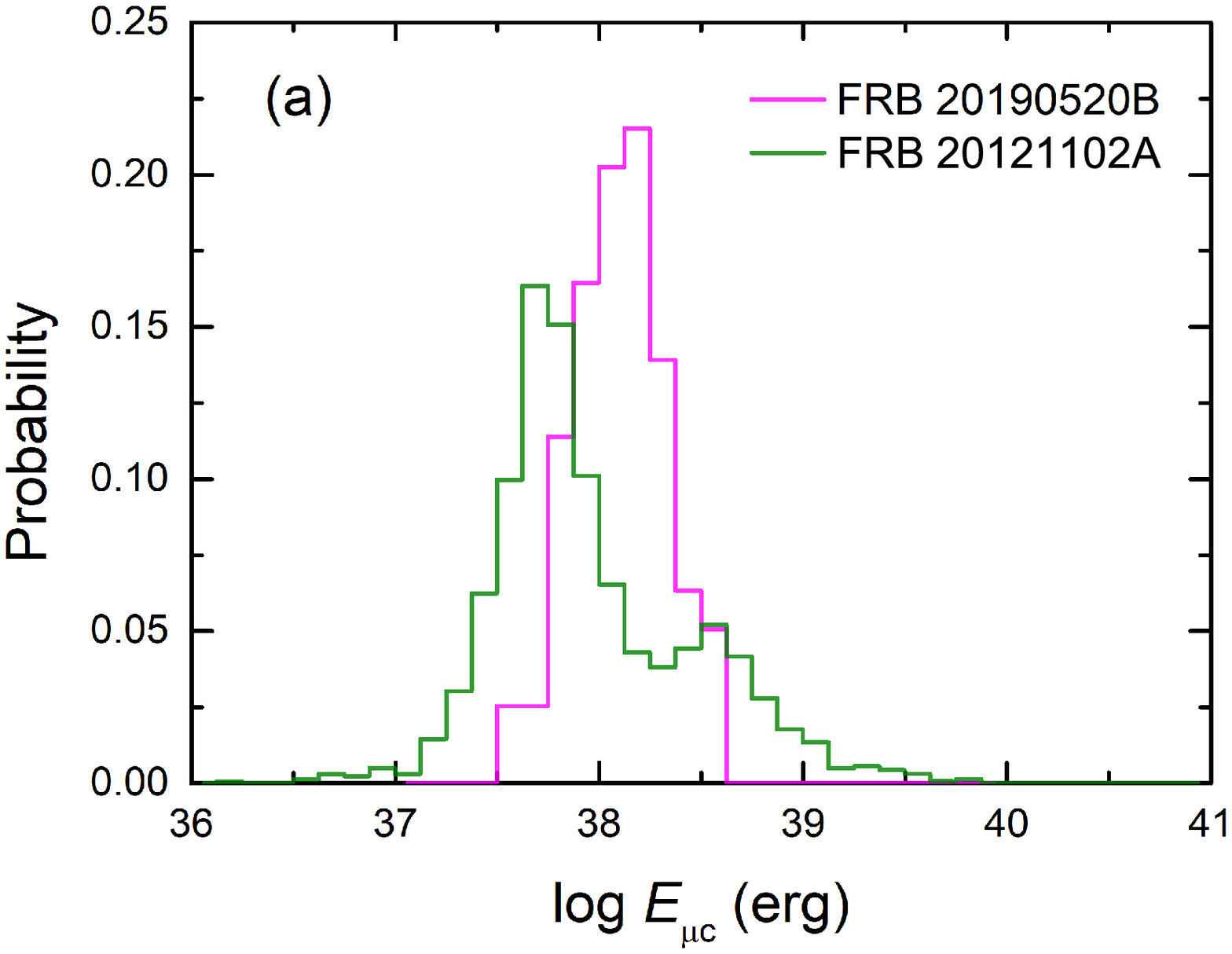}\hspace{-0.1in}
\includegraphics[width=0.45\linewidth]{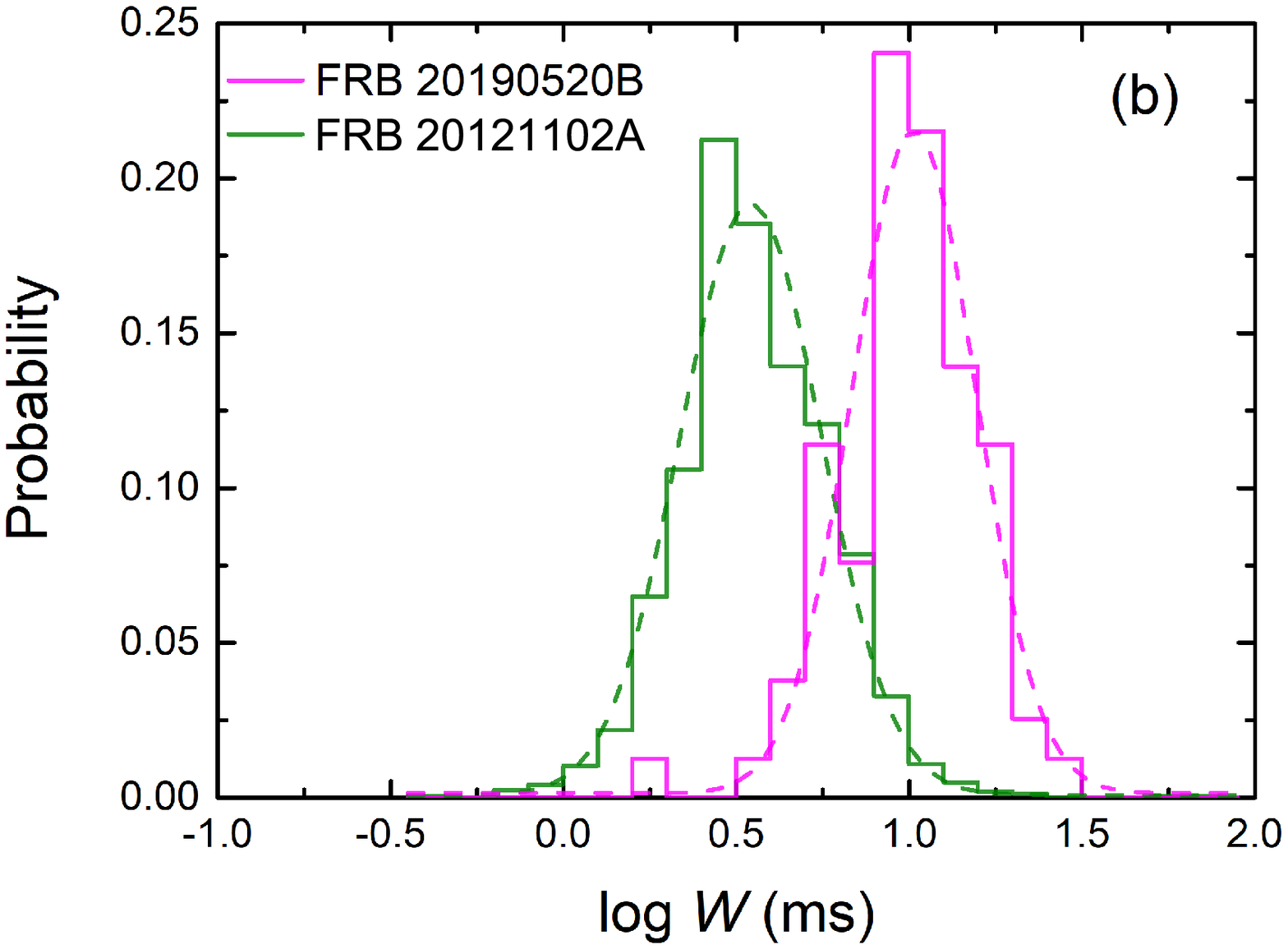}\hspace{-0.1in}
\caption{Histograms of the burst specific energy $E_{\rm \mu_{\rm c}}$ at the instrumental central frequency $\mu_{\rm c}$ and the burst duration in the source rest frame ($W$) of FRB 20190520B (magenta) in comparison with FRB 20121102A (green) detected by FAST. The dashed lines in panel(b) mark the Gaussian fit to the distributions.
\label{fig:obs}}
\vspace{-0.5cm}
\end{figure}

 Figure \ref{fig:obs} shows the $E_{\rm \mu_{\rm c}}$ and $W$ distributions of FRB 20190520B in comparison with that of FRB 20121102A observed with the FAST telescope. It is found that $E_{\rm \mu_{\rm c}}$ ranges from $3.6\times 10^{37}$ to $\sim 4.04\times 10^{38}$ erg, being narrower than that of FRB 20121102A ([$4\times 10^{36}, 8 \times 10^{39}]\,\mathrm{erg}$). It peaks at $\sim 2 \times10^{38}$ erg, which is roughly the separation of the bimodal $E_{\rm \mu_{\rm c}}$ distribution of FRB 20121102A \citep{2021Natur.598..267L}.  The $\log W$ distributions of FRB 20190520B and FRB 20121102A can be fitted with a Gaussian function, 
$p(\log W; \log W_c,\sigma_{\log W})
\propto \frac{1}{\sigma_{\log W}} e^{-(\log W-\log W_c)^{2}/2\sigma_{\log W}^{2}}$. 
We obtain $\log W_c/ \mathrm{ms}=1.03$ ($W_c=11$ ms) and $\sigma_{\log W}=0.18$ with $R^2=0.96$ for FRB 20190520B and $\log W_c/ \mathrm{ms}=0.55$ ($W_c=3.55$ ms) and $\sigma_{\log W_c}=0.20$ with $R^2=0.97$ for FRB 20121102A, where $R^2$ is the $R^2$ for measuring the goodness of the fits\footnote{$R^2$ is calculated as $R^2=1-RSS/TSS$, where $TSS$ is the total sum of square of the data and $RSS$ is the sum of the squares of residuals, i.e. deviations of the model from the data. $R^2$ is a value between 0 and 1. A $R^2$ value closer to 1 indicates a better fit.}. The $W_c$ value of FRB 20190520B ($10.7$ ms) is 3 times of the $W_c$ of FRB 20121102A ($3.5$ ms). 
 
 Note that the 79 bursts from FRB 20190520B selected for the above analysis were observed with the FAST telescope by adopting a detection threshold of 0.009 Jy ms \citep{2022Natur.606..873N}, but the selected 1652 bursts of FRB 20121102A were observed with the FAST telescope by using a threshold of 0.015 Jy ms \citep{2021Natur.598..267L}. The redshifts of the two FRBs are also different. We suspect whether the differences of the observed $E_{\rm \mu_{\rm c}}$ and $W$ distributions between two FRBs are caused by the different selection thresholds and different distances. Assuming that FRB 20121102A is located at the redshift of FRB 20190520B ($z=0.241$), we calculate the fluences of its bursts and then select the bursts above the $90 \%$ complete fluence threshold of FRB 20190520B (0.023 Jy ms). We find that the detectable bursts are reduced from 1652 to 1286 and 366 low-energy bursts are ruled out of the sample. However, the bimodal $E_{\rm \mu_{\rm c}}$ distribution still exists and no significant statistical change in the probability distribution of $W$ is found. Therefore, the differences of the observed $E_{\rm \mu_{\rm c}}$ and $W$ distributions between the two FRBs are not due to the observational selection effects, and probably are intrinsically different.

\subsection{Comparison of the Peak Frequency and Spectral Width with Different Telescope Observations}

\begin{figure}
\centering
\includegraphics[width=0.45\linewidth]{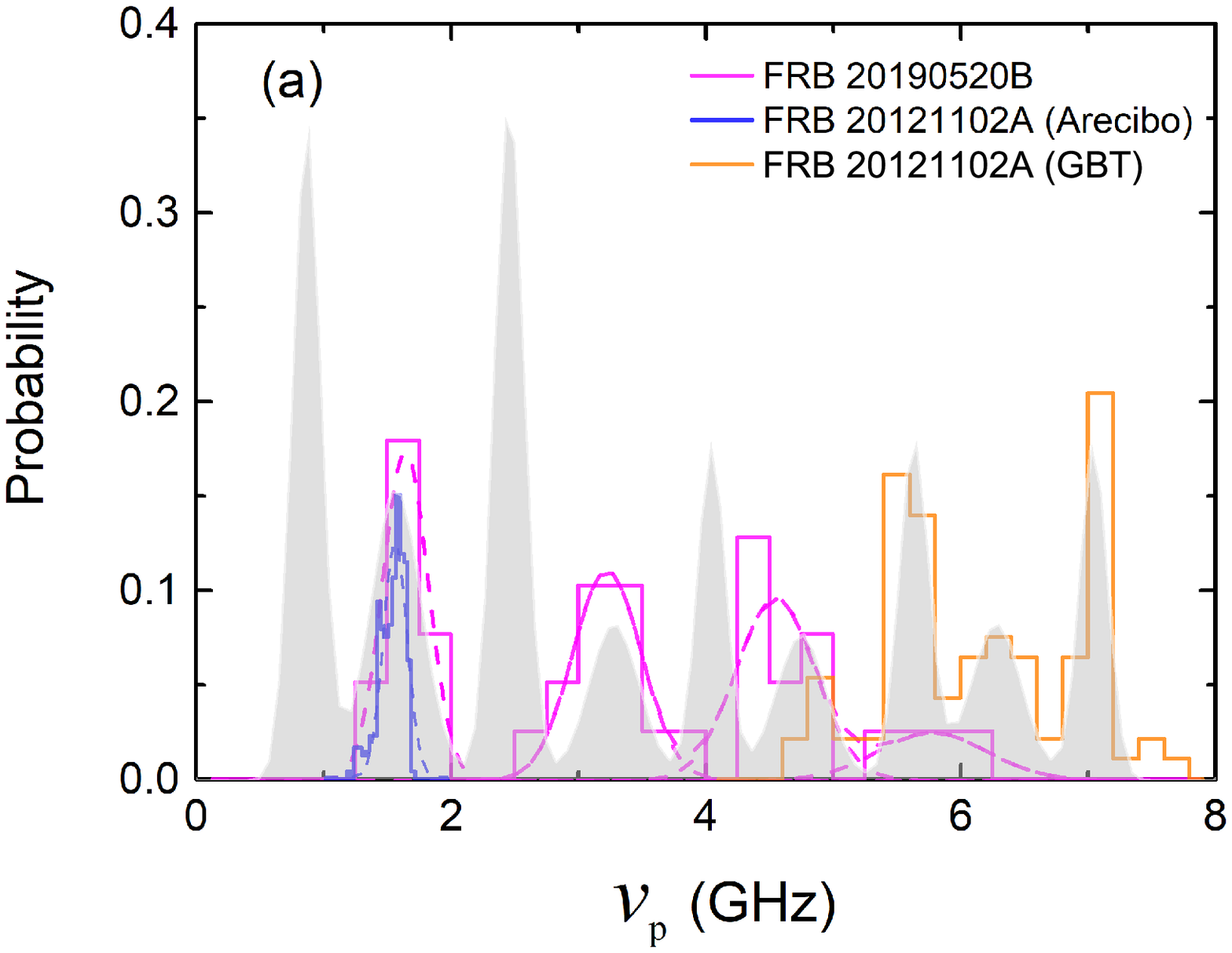}\vspace{-0.1in}
\includegraphics[width=0.45\linewidth]{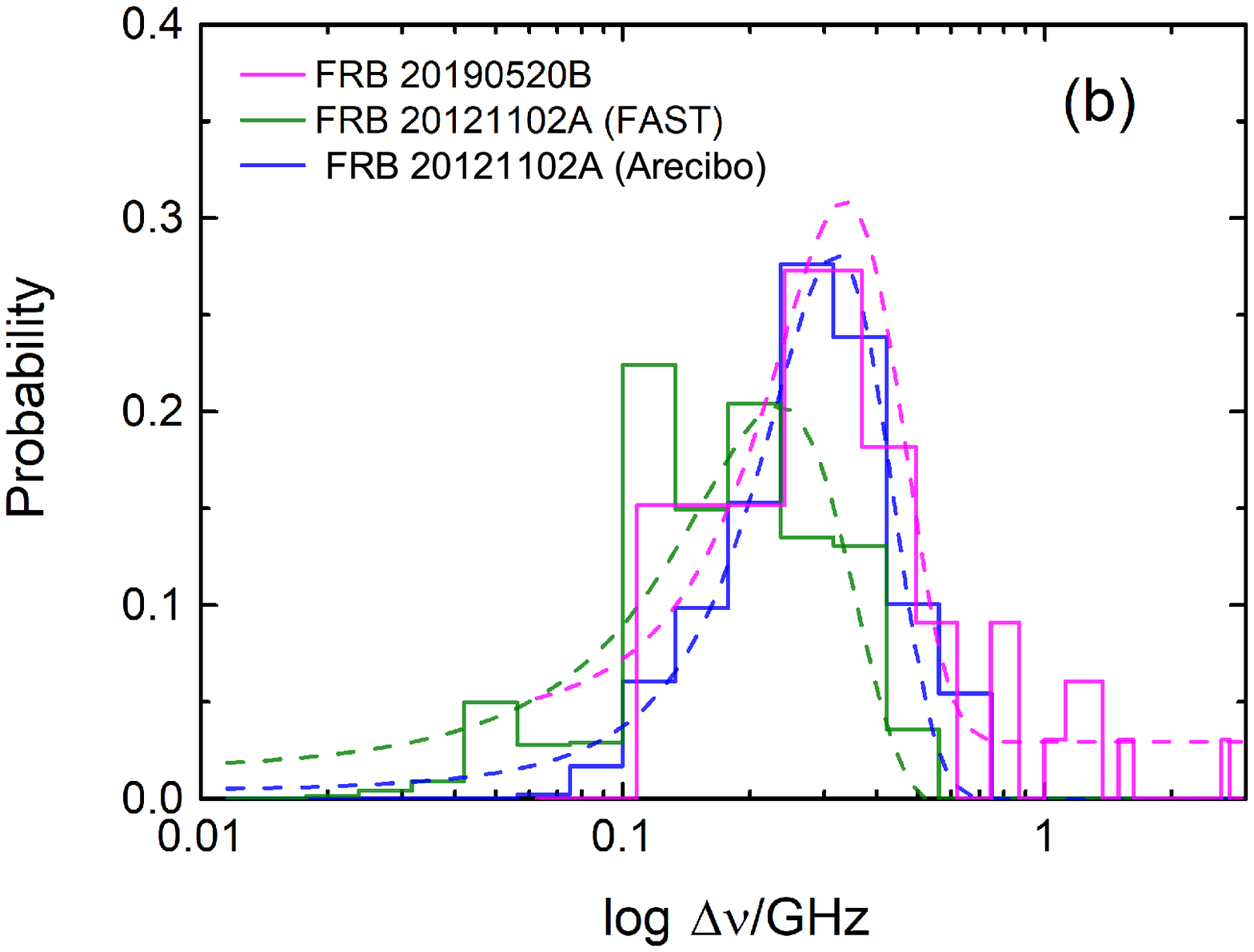}\hspace{-0.1in}
\caption{Distributions of the radiating peak frequency $\nu_{p}$ and the burst emission bandwidth $\log \Delta \nu$ of FRB 20190520B (magenta) in comparison with that of FRB 20121102A derived from \protect\cite{2022ApJ...941..127L}. The dashed lines represent the Gaussian function fits.}
And the shaded area in panel(a) marks the constructed $\nu_{p}$-distribution of FRB 20121102A.
\label{fig:obs_test}
\vspace{-0.2cm}
\end{figure}

We investigate the burst observed spectral properties (peak frequency $\nu_{p}$ and spectral width $\Delta\nu$) of FRB 20190520B with the data reported in Table \ref{collect}. These broadband data have a frequency coverage from 1-8 GHZ. They are observed with telescopes in different monitoring campaigns. Figure \ref{fig:obs_test}(a) shows the observed $\nu_{p}$-distribution of FRB 20190520B in comparison with FRB 20121102A. Similar to the $\nu_p$ fringe pattern of FRB 20121102A reported by \cite{2022ApJ...941..127L}, the $\nu_{p}$-distribution of FRB 20190520B demonstrates some peaks. The most striking ones are the two peaks in 1-4 GHz. Tentative peaks at 4-5 GHz and 5.5-6.5 GHz are also observed in the $\nu_p$ probability distribution. We fit each peak with a Gaussian function, %
\begin{equation}
p=\sum_{i=1}^{2}\frac{p_i}{\sigma_{\nu_{p,i}}\sqrt{2\pi}}e^{-\frac{(\nu_{\rm p,i}-\nu_{\rm p,c,i})^2}{2\sigma_{\nu_{p,i}}^2}},
\label{nu_p}
\end{equation}   
where $p_i$, $\nu_{\rm p,c,i}$, and $\sigma_{\nu_{p,i}}$ are the peak amplitude of the Gaussian function, the center value of $\nu_p$ and the standard deviation of the $i$th Gaussian function (i=1, 2, 3, 4). 
The fitting results are reported in Table \ref{table}, the $R^2$ of the fit is 0.88, and the fitting curve is illustrated in Figure \ref{fig:obs_test}(a). 

Comparing this fringe pattern with that of FRB 20121102A \citep{2022ApJ...941..127L}, we observe that the first peak in the $L$-band (1-2 GHz), is almost overlapped with the $\nu_p$ distribution peak of FRB 20121102A observed with the Arecibo telescope. The second peak is in the $S$-band (2.5-4 GHz). No burst observed $\nu_p$ in this frequency range of FRB 20121102A is available, but a peak at $3\sim 3.3$ GHz in the $\nu_p$ distribution of FRB 20121102A is inferred by extrapolating the $\nu_p$ distribution of the GBT data to the low-frequency end \citep{2022ApJ...941..127L}. This peak is consistent with the observed $\nu_p$ distribution peak of FRB 20190520B in 2.5-4 GHz. In the $C$-band (4-8 GHz), tentative peaks are also found. Compared with FRB 20121102A, the number of bursts in FRB 20190520B seems to be reduced when the observing frequency is beyond 5 GHz.

\begin{table}
    \centering
\caption{The Intrinsic $\nu_{\rm p}$ Distribution of FRB 20190520B in the Frequency Coverage [1, 6.2] GHz Inferred from the Observed $\nu_{\rm p}$ Distributions with the VLA, GBT, and Parkes Observations.}\label{table}
\begin{tabular}{|c|c|c|c|}
\hline
$i$th fringe & ${p_{\rm i}}$ & $\nu_{\rm p,c,i}$&$\sigma_{\nu_p,i}$\\
\hline
1& 0.078 &1.64& 0.18 \\
2 &0.075 & 3.23&0.27\\
3&  0.075 & 4.55&0.31\\
4 & 0.027 & 5.78&0.43\\
\hline
\end{tabular}\\
\end{table}

Figure \ref{fig:obs_test} (b) shows the observed $\Delta\nu$ distribution of FRB 20190520B in comparison with the bursts of FRB 20121102A observed with the FAST and Arecibo telescopes. It is found that the range of the $\Delta\nu$ distribution of FRB 20190520B is much broader than that of FRB 20121102A. It extends up to 2.6 GHz. Fitting the $\log \Delta \nu$ distribution of FRB 20190520B with a Gaussian function $N(\log \Delta\nu; \log \Delta \nu_c,\sigma_{\log \Delta \nu})$, we obtain $\log \Delta\nu_c=-0.45$ GHz and $\sigma_{\log \Delta \nu}=0.20$. The $R^{2}$ of the fit is 0.74. For FRB 20121102A, we derive $\log \Delta\nu_c=-0.77$ GHz and $\sigma_{\log \Delta \nu}=0.28$ ($R^{2}$=0.93) from the observations of the FAST telescope \citep{2021Natur.598..267L} and $\log \Delta\nu_c/{\rm GHz}=-0.54$ and $\sigma_{\log \Delta \nu}=0.18$ ($R^{2}$=0.98) from the observations of the Arecibo telescope \citep{2022MNRAS.515.3577H}. Excluding those bursts that have a $\Delta \nu> 1$ GHz, the $\log \Delta \nu$ distribution of FRB 20190520B is roughly consistent with that of FRB 20121102A observed with the Arecibo telescope. 

The $\Delta\nu$ of FRB 20121102A observed with the FAST telescope averagely is smaller than that observed with the Arecibo telescope. Observation in broad frequency coverage is critical to measure the real spectral width. The $\Delta\nu$ distribution of FRB 20190520B is generated by using the broadband observations with the GBT and VLA.   
We should note that the $\Delta \nu$ values of 6 bursts observed with the GBT at $C$-band (4-8 GHz) of FRB 20190520B are larger than 1 GHz, and the $\Delta \nu$ of 1 burst is even 2.63 GHz.  It is unclear whether the extremely large  $\Delta \nu$ is intrinsic or due to
insufficiently resolving overlapping bursts with low spectral extent. Searching an FRB pulse signal over a large bandwidth is insufficient to distinguish the weak bursts with low spectral coverage, and a narrow band search might increase the detection rate.  
For example, 21 bursts were found by conducting a full band dispersion search across 4 GHz full-band observation of FRB 20121102A with the GBT telescope in the $C$-band (4-8 GHz) \citep{2018ApJ...863....2G}. However, 72 new bursts were discovered through the narrow bandwidth search using a neural network machine learning algorithm \citep{2018ApJ...866..149Z}. The substantial increase in detection rate results from resolving bursts with low spectral occupancy in broadband. In \cite{2021RNAAS...5...17F}, 8 more new bursts were found by dividing the entire band into 8 sub-bands (each spanning $\sim$ 500 MHz) using the previous GBT observation reported by \cite{2018ApJ...863....2G}. Moreover, \cite{2022arXiv220211112A} employed a sub-banded search by dividing the total bandwidth of GBT at $C$-band (4-8 GHz) into sub-bands of bandwidth 0.75 GHz and 1.5 GHz, and 16 bursts were detected from the $C$-band observation for FRB 20190520B. Therefore, those bursts with large $\Delta \nu$ may be caused by the searching algorithm for FRB 20190520B.

\section{Monte Carlo Simulation Analysis}

\label{sec:Assumptions}
 Our above analysis presents the $\nu_p$ distribution in $\sim 1-6$ GHz and the $\Delta \nu$ distributions of FRB 20190520B with data observed by the GBT, VLA, and Parkes telescopes. To analyze if the broadband spectral features seen for FRB 20190520B could be intrinsic, we investigate its intrinsic energy function via Monte Carlo simulation by comparing the simulated and observed $E_{\rm \mu_{\rm c}}$ distributions of the FAST sample. Our simulation analysis strategy is the same as that presented in \cite{2022ApJ...941..127L}. 

\subsection{Spectral model}
The intrinsic spectrum of FRB 20190520B is assumed to be a Gaussian profile,  
\begin{equation}
F_\nu
=\frac{F^{\rm int}}{\sigma_{\rm s} \sqrt{2 \pi}} \exp \left[\frac{-\left(\nu-\nu_{p}\right)^{2}}{2\sigma_{\rm s}^{2}}\right], \label{Gauss}
\end{equation}
where $F^{\rm int}$ ($F^{\rm int}=E/4\pi D_L^2$) is the intrinsic fluence of a burst having energy $E$, $\nu_{p}$ and $\sigma_{\rm s}$ are the peak frequency and 1$\sigma$ width of the Gaussian spectrum, respectively. The $\sigma_s$ value is taken as $\Delta \nu^{\rm sim}/2$, and $\Delta \nu^{\rm sim}$ is bootstrapped from the observed probability distribution of $\Delta \nu$ of FRB 20190520B,  i.e., $p(\rm log \Delta \nu)\propto 10^{(\log \Delta \nu -\log \Delta \nu_c)^2/2\sigma_{\log \Delta \nu}^2}$, where $\log \Delta \nu_c=-0.46$ and $\log \Delta \nu=0.09$. Since the bandpass of FAST is 1.05-1.45 GHz and the observed spectral width is usually $\sigma_s<1$ GHz, the $\nu_p$ values of most detectable bursts should be in 1-2 GHz, and occasionally above 2 GHz. Thus, we bootstrap the $\nu_p$ values from the $\nu_p$ probability distribution of Eq. (\ref{nu_p}) by considering only the two peaks in the range of 1-4 GHz for simulating the FAST observations. 

We define the burst radiating frequency range as $[\nu_1,\nu_2]$, where $\nu_1=\nu_{p}-\sigma_{\rm s}$, and $\nu_2=\nu_{p}+\sigma_{\rm s}$. The observable frequency range $[\lambda_1,\lambda_2]$ with the FAST telescope in the bandpass $[\mu_1,\mu_2]=[1.05, 1.45]$ is given by $\lambda_1=\max(\nu_1, \mu_1)$ and $\lambda_2=\min(\nu_2, \mu_2)$. This ensures that $[\lambda_1, \lambda_2]$ is within the FAST telescope bandpass. The ``statistical'' spectral slope $\beta^{\rm sim}$ is used to represent the global spectral shape (rising, decaying, or flattening spectrum) in the observable frequency coverage \citep{2022ApJ...941..127L}. 

 A mock burst is characterized by a parameter set of $\{E,\nu_p, \sigma_s\}$ by setting an intrinsic energy function. For a mock burst with the intrinsic energy $E$, we calculate its specific fluence in the given frequency range $[\lambda_1, \lambda_2]$, screen it with the detection threshold $F_{\rm th}=9$ mJy ms of the FAST telescope for selecting the bursts of FRB 20190520B \citep{2022Natur.606..873N} and calculate its $E_{\rm \mu_{\rm c}}$ values using Eq (\ref{Energy}).
\subsection{Energy Function Model: Power-Law Function}
\begin{figure}
\centering
  \includegraphics[width=0.5\linewidth]{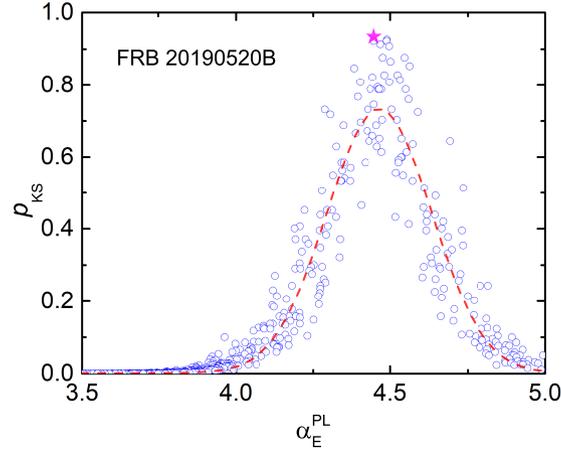}
\caption{The $p_{\mathrm{KS}}$ as a function of $\alpha_{E}^{\rm PL}$ for FRB 20190520B derived from the simulation by modeling the intrinsic $E$-distribution as a simple power-law function. The red dotted line indicates its Gaussian profile. The magenta star marks the maximum likelihood parameter ($\alpha^{\rm PL}_E=4.46$ and $p_{\rm {KS}}=0.93$).}\label{fig:PL}
\vspace{-0.1cm}
\end{figure}

\begin{figure}
\includegraphics[width=0.45\linewidth]{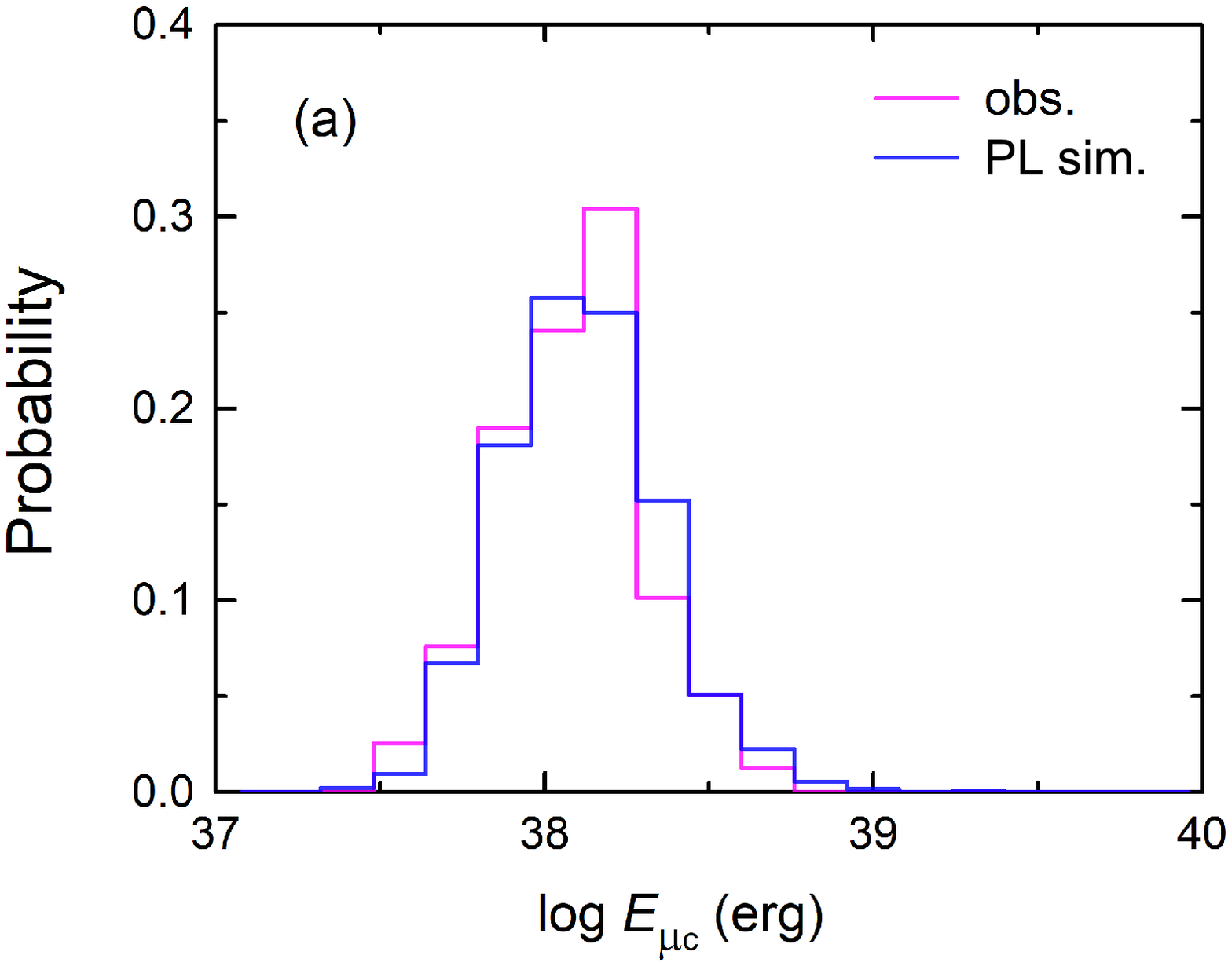}
\includegraphics[width=0.45\linewidth]{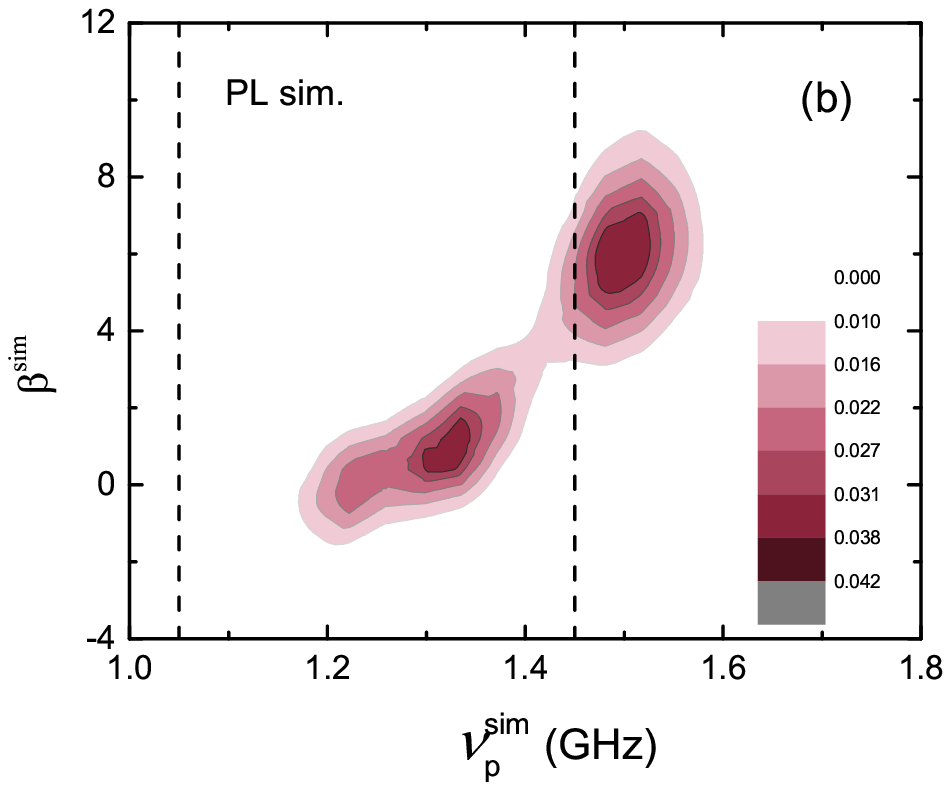}
\caption{Comparison of the simulated (blue) and observed (pink) $E_{\rm \mu_{\rm c}}$-distributions in the PL energy function scenario of FRB 20190520B by adopting $\alpha^{\rm PL}_E=4.46$ (panel a), together with the probability contours of the simulated sample in the $\nu_{p}^{\rm sim}-\beta^{\rm sim}$ plane (panel b), where the dashed lines mark the FAST bandpass (1.05-1.45 GHz). 
\label{fig:PL_new_best}}
\vspace{-0.1cm}
\end{figure}

We first assume the intrinsic energy differential distribution ($E-$distribution) of FRB 20190520B over the range of $[10^{37},10^{42}]$ erg as a simple power-law (PL) function,  
\begin{equation}
    \frac{dp(E)}{dE}\propto E^{-\alpha^{\rm PL}_E}\label{PL}, 
\end{equation}
where $\alpha^{\rm PL}_E$ is randomly picked up from a uniform distribution in the range of [1.1, 5]. For a given $\alpha^{\rm PL}_E$, we generate a mock sample of 2000 detectable bursts and measure the maximum likelihood of $\alpha^{\rm PL}_E$ with the probability of the K-S test ($p_{\rm KS}$)\footnote{The Kolmogorov-Smirnov test (K-S test) is a nonparametric statistical hypothesis method which is used to assess whether two groups come from the same population. One is likely to reject the null hypothesis, i.e. two samples from the same distribution. Two samples hold an identical distribution at a 5$\%$ level of significance when the p-value of the test statistic is less than 0.05. Here the two groups are claimed to follow the same distribution if $P_{\rm KS}>10^{-4}$ with a confidence level of $>3\sigma$.} for the observed and mock $E_{\rm \mu_{\rm c}}$ distribution in FAST sample in \cite{2022Natur.606..873N}. Figure \ref{fig:PL} displays $p_{\mathrm{KS}}$ as a function of $\alpha^{\rm PL}_E$.
The profile of the curve can be globally described with a Gaussian function peaking at $\alpha^{\rm PL}_ E=4.46\pm 0.17$ with the maximum $p_{\rm KS}$ of 0.93. Taking $\alpha^{\rm PL}_ E=4.46$, the comparison of the observed and mock $E_{\rm \mu_{\rm c}}$ distributions, together with the corresponding $\nu^{\rm sim}_p$ and $\beta^{\rm sim}$ distributions, are shown in Figure \ref{fig:PL_new_best}. One can find that the observed $E_{\rm \mu_{\rm c}}$ can be reproduced by the simulation.

 We also show the probability contours of the simulated sample with the best parameter set in the $\nu^{\rm sim}_p$-$\beta^{\rm sim}$ plane. Two distinct peaks are illustrated. 
The $\nu^{\rm sim}_p$ of the first peak is in the FAST bandpass, centering at around the middle of the FAST bandpass, i.e. $\sim 1.35$ GHz. The other $\nu^{\rm sim}_p$ peak is at $\sim 1.5$ GHz, close to the $\nu_p$ peak in the range of 1-2 GHz. Such a $\nu^{\rm sim}_p$ bimodal distribution results from the selection effect of the FAST telescope. For those bursts that have a $\nu_p$ closer to the middle of the bandpass, the FAST telescope observes the main emission part of these bursts, leading to a higher detection probability. This selection effect makes the $\nu^{\rm sim}$ distribution peak around $1.35$ GHz. The spectral shape of these bursts is relatively flat, i.e. $\beta^{\rm sim}\leq4$. For the $\nu_p^{\rm sim}$ peak at 1.5GHz, the FAST telescope observes only the rising parts of the Gaussian spectra of these bursts, making the bursts have a relatively steep spectral shape, i.e. $\beta^{\rm sim}>4$.

\subsection{Energy Function Model: Cut-Off Power-Law Function}

\begin{figure}
\includegraphics[width=0.45\linewidth]{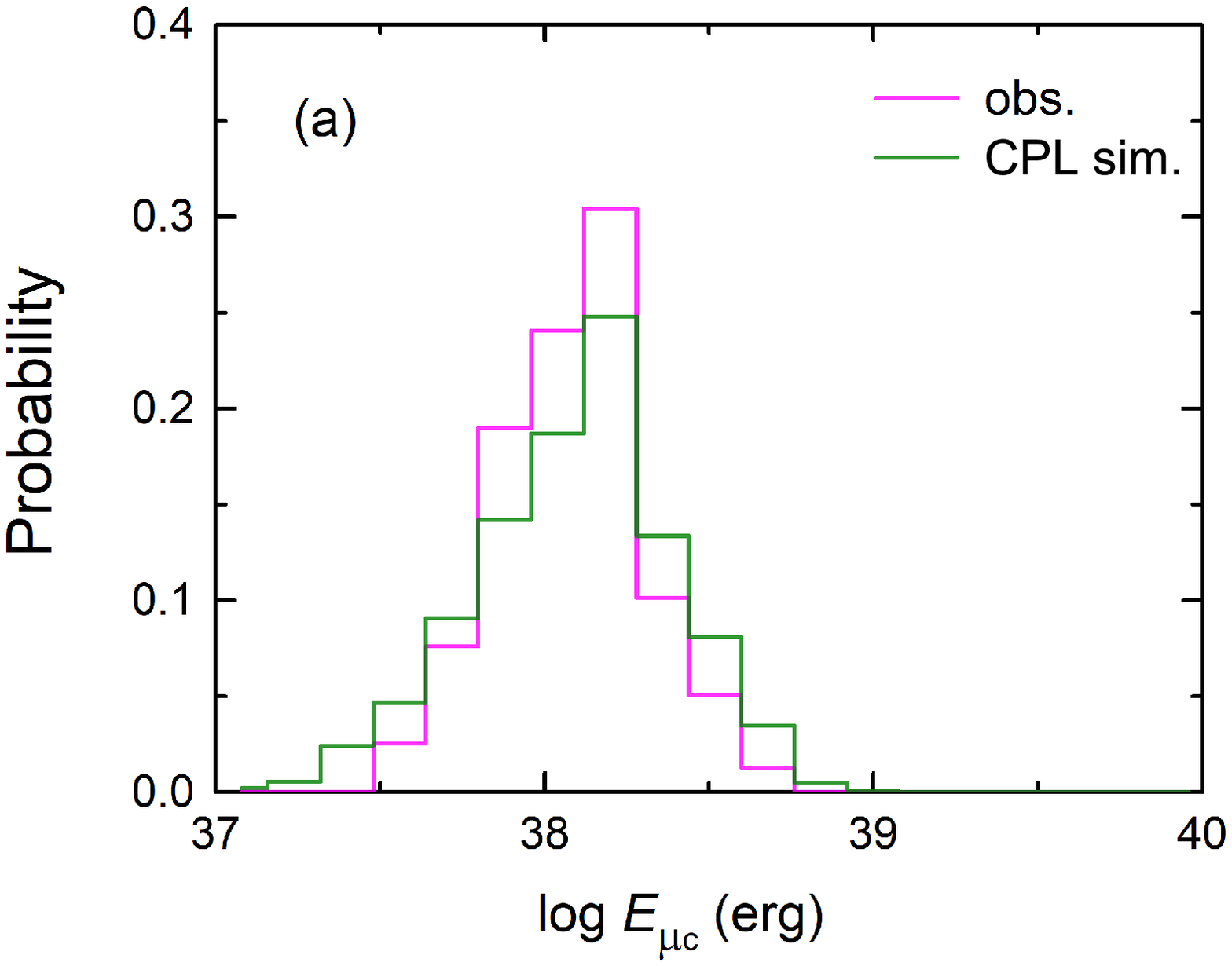}
\includegraphics[width=0.45\linewidth]{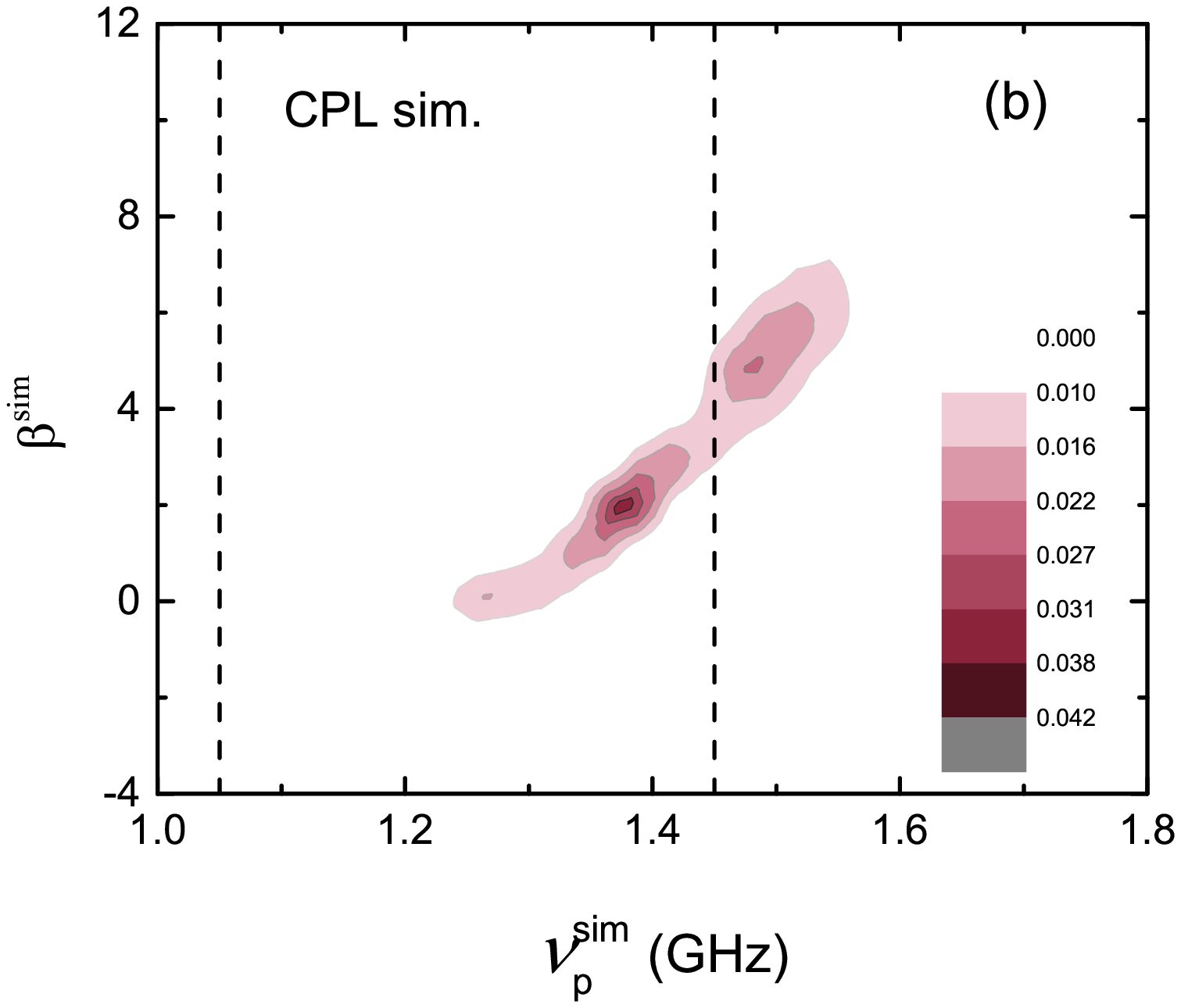}
\caption{Comparison of the simulated (green) and observed (pink) $E_{\rm \mu_{\rm c}}$ distributions in the CPL energy function scenario of FRB 20190520B by adopting $\alpha^{\rm CPL}_E=0.46$ and $\log E_c/{\rm erg}=37.87$ (panel a), together with the probability contours of the simulated sample in the $\nu_{p}^{\rm sim}-\beta^{\rm sim}$ plane (panel b), where the dashed lines mark the FAST bandpass (1.05-1.45 GHz).
\label{fig:CPL_best}}
\vspace{-0.1cm}
\end{figure}

The extremely steep slope of the derived energy function above motivates us to explore the possible results by adopting a cut-off power-law (CPL) energy function, i.e. 
\begin{equation}
\frac{dp(E)}{dE}\propto E^{-\alpha^{\rm CPL}_E} e^{-E/E_c} \label{CPL},   
\end{equation}
where $\alpha^{\rm CPL}_{E}$ is the power-law index and $E_c$ is the cutoff energy. We assume that $\alpha_E^{\rm CPL}$ and $\log E_c/{\rm erg}$ are uniformly distributed in the ranges of [-3, 4] and [$37.5, 39]$, respectively. We randomly pick up a set of $\{\alpha_E^{\rm CPL}$ $\log E_c\}$ for constructing an energy function. We make simulations following the same approach as that for the PL model. Figure \ref{fig:PL_new_best} shows the $\log P_{\rm KS}$ contours in the $\alpha_E^{\rm CPL}-\log E_c$ plane. Adopting $p_{\rm KS}>10^{-4}$, $\alpha_E^{\rm CPL}$ is constrained as $-0.81\le \alpha^{\rm CPL}_E\le 3.06$.  $\log E_c$ is poorly constrained with the criterion of $p_{\rm KS}=10^{-4}$. With a criterion of $p_{\rm KS}=10^{-1}$, we have $\log E_{\rm c}/{\rm erg}\in [37.5,38.0]$. The parameter set $\{\alpha_E^{\rm CPL}, \log E_c/{\rm erg}\}=\{0.47, 37.87\}$ has the maximum likelihood, i.e. $p_{\mathrm{KS}}$=0.60. Figure \ref{fig:CPL_best} shows the comparison of the observed and simulated $E_{\rm \mu_{\rm c}}$ distributions, together with the corresponding  probability contours of the simulated sample in the $\nu^{\rm sim}_p$-$\beta^{\rm sim}$ plane. The observed $E_{\rm \mu_c}$ distribution is also well reproduced with the CPL model. 

\begin{figure}
\centering
\includegraphics[width=0.45\linewidth]{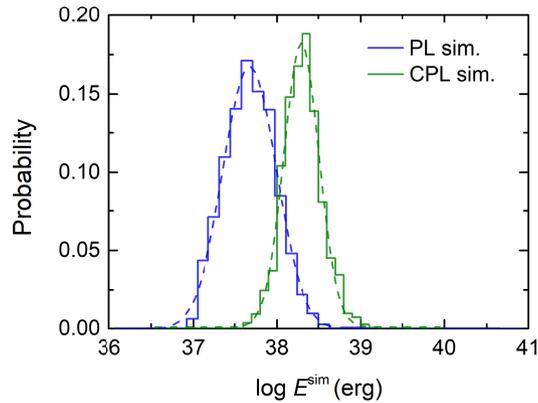}\hspace{-0.1in}
\caption{Intrinsic energy Distributions of the simulated sample derived from the PL (blue) and the CPL models (green) for FRB 20190520B.
\label{fig:Eint}}
\vspace{-0.2cm}
\end{figure}

 It is interesting that the probability contours of the simulated sample in the $\nu^{\rm sim}_p$-$\beta^{\rm sim}$ plane are significantly different from that derived from the PL energy function model. The clear bimodal feature in the $\nu_p^{\rm sim }$ and $\beta^{\rm sim}$ distributions derived from the PL energy function model is smeared out in the results from the CPL model.
Since the derived PL energy function is extremely steep. i.e. $\alpha^{\rm PL}_{\rm E}=4.46\pm 0.17$, the probability of bursts at the low energy end is extremely large with respect to that of the CPL energy function model. Note that low energy bursts could be detectable if their $\nu_p$ are close to the middle of the FAST bandpass. Thus, the fraction of low energy bursts with a $\nu_p$ value within the FAST bandpass derived from the PL function model is larger than that of the CPL model. This leads to not only a more clear bimodal signature in the $\nu^{\rm sim}_p$ and $\beta^{\rm sim}$ distributions but also a smaller average intrinsic bolometric burst energy ($E^{\rm sim }$) for the PL model than that for the CPL model. We compare the $E^{\rm sim}$ distributions of the simulated samples in Figure \ref{fig:Eint}. 
It is found that the peak energy of the $\log E^{\rm sim}$ distribution derived from the CPL model is $2.24\times 10^{38}$ ergs, being larger than the peak of the $\log E^{\rm sim}$ distribution derived from the PL model by a factor of 5. As shown in Figure \ref{nu_p}, we do not find any bimodal feature in the observed $\nu_p$ distribution in 1-2 GHz. Therefore, the CPL energy function model is preferred than the PL model.   

\begin{figure}
\centering
\includegraphics[width=0.45\linewidth]{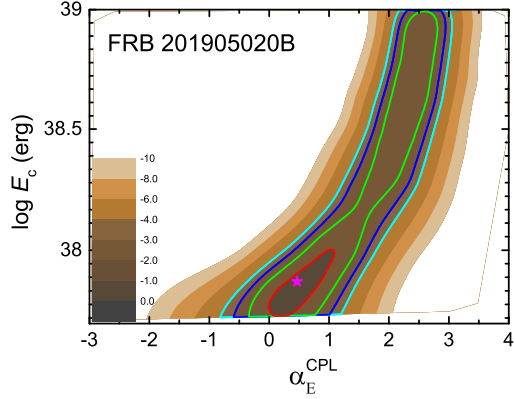}\hspace{-0.5in} 
\caption{The $\log P_{\mathrm{KS}}$ contours in the $\log E_c-\alpha_{E}^{\rm CPL}$ plane for FRB 20190520B derived from the simulation by modeling the intrinsic $E$-distribution as a cut-off power-law (CPL) function. The cyan, blue, green, and red lines mark the contours of $\log p_{\mathrm{KS}}$=-4, -3, -2, and -1, respectively. The magenta star marks the maximum likelihood ($p_{\mathrm{KS}}=0.60$) parameter set $\{\alpha^{\rm CPL}_E=0.47, \log E_c/{\rm erg}=37.87\}$.
\label{fig:CPL_counter}}
\vspace{-0.5cm}
\end{figure}

\section{Summary and Discussion} \label{sec:Discussion}
We have presented a comparative analysis of FRB 20190520B with FRB 20121102A by compiling a sample of bursts observed with observations in multiple wavelengths. It is found that the $E_{\rm \mu_{\rm c}}$ distribution of FRB 20190520B observed with the FAST telescope in the $L$-band (1.05-1.45 GHz) spans only one order of magnitude ($4\times 10^{37}\sim 10^{38}$ ergs), which is around the median of the bimodal $E_{\rm \mu_{\rm c}}$ distribution of FRB 20121102A. The burst durations of the two FRBs observed with the FAST telescope can be fitted with a log-normal distribution, which peaks at $\sim 10.72$ ms 
for FRB 20190520B and peak at $\sim$3.54 ms 
for FRB 20121102A.
The $\nu_p$ distribution of the bursts of FRB 20190520B illustrates a discrete pattern and the radiation spectrum is narrow-banded ($\log \Delta\nu=-0.45\pm 0.20$ GHz), similar to that observed in FRB 20121102A \citep{2022ApJ...941..127L}.  

Based on the observed $\nu_p$ and $\Delta \nu$ distributions, we describe the radiating spectrum as a Gaussian profile and make a Monte Carlo simulation analysis. Our results suggest that the PL function model with $\alpha^{\rm PL}_E=4.46\pm 0.17$ and the CPL energy function model with $\alpha^{\rm CPL}_E=0.46$ and $\log E_c/{\rm erg}=37.87$ can comparably reproduce the observed $E_{\rm \mu_{\rm c}}$ distribution of FRB 20190520B with the FAST telescope. The two model predicted different distributions of $\nu^{\rm sim}_p$ and $\beta^{\rm sim}$.
A clear bimodal signature is seen in the $\nu^{\rm sim}_p$ distribution in 1-2 GHz and in the $\beta^{\rm sim}$ distribution of the simulated sample derived from the PL model, but this feature is smeared out in the scenario of the CPL model. Both the PL and CPL models can represent the observed $E_{\mu_c}$ distribution. The $\nu^{sim}_p$ distribution could provide some clues to the energy function models. Because no bimodal feature is found in the observed $\nu_p$ distribution in 1-2 GHz, the clear bimodal distribution of $\nu^{\rm sim}_p$ disfavors the PL model and the CPL energy function model is a preferred one. In comparison with FRB 20121102A, a PL energy function model with $\alpha_{E}=1.82^{+0.10}_{-0.30}$ can well reproduce the $E_{\rm \mu_{\rm c}}$ distribution of FRB 2021102A observed with the FAST telescope \citep{2022ApJ...941..127L}. These results indicate that the energy function of FRB 20190520B is intrinsically different from FRB 20121102A.

As shown in our analysis, $\nu_p$ and $\beta$ distributions are critical for examining the energy function model. Our current analysis with data observed with different telescopes may be significantly affected by observational biases. The $\nu_p$ distributions of FRB 20121102A and FRB 20190520B are derived from bursts observed with different telescopes. They could be biased by the observing bands of the respective telescopes. The emission edges of a large fraction of bursts observed with the FAST telescope extend beyond the observed bandpass (e.g. \citealt{2021ApJ...920L..18A,2022Natur.606..873N,2022ApJ...941..127L}). We lost the spectral information of these bursts. For avoiding the narrow-band bias, we use only the broadband data observed with GBT, VLA, and Parkes telescopes to generate the $\nu_p$ and $\Delta \nu$ distribution in this analysis. However, the samples are still very small and they are collected from observations with different telescopes with different detection thresholds at different observing bands. Our conclusions are potentially affected by observational biases. More bursts in broadband observations are critical for examining our conclusions.   

\section*{acknowledgments}
We very much appreciate thoughtful and constructive comments and suggestions from the anonymous referee. We also thank Chen-Hui Niu, Pei Wang, and Ji-Gui Cheng for their helpful discussion. We acknowledge the use of the public data from the FAST/FRB Key Project. F.L. is supported by the Shanghai Post-doctoral Excellence Program and the National SKA Program of China (grant No. 2022SKA0130103). E.W.L is supported by the National Natural Science Foundation of China (grant No.~12133003).

\section*{Data Availability}

The data underlying this article will be shared on reasonable request to the corresponding authors.











\newpage
\begin{longtable}[!htbp]{ccccc}
\caption{
The temporal and spectral properties of FRB 20190520B observed With various radio telescopes.} 
\label{collect}\\
\hline
\hline
burst ID&$\omega$ &$\nu_{p}$ &$\Delta \nu$&$E_{\mu \rm c}$\\
--&(ms)&(GHz) &(GHz)&$\left(\times 10^{37} \mathrm{erg}\right)$\\
\hline								
\hline			
\textbf{FAST}\footnote{L-band (1.05-1.45 GHz)}&&\cite{2022Natur.606..873N}&&\\
P1&8.7(8)&--&--&9.1(0)\\
P2&16.4(6)&--&--&3.6(0)\\
P3&7.1(8)&--&--&6.4(0)\\
P4&7.3(10)&--&--&6.2(0)\\
P5&10.4(2)&--&--&8.0(0)\\
P6&10.2(17)&--&--&6.4(0)\\
P7&20.3(4)&--&--&13.0(2)\\
P8&22.8(83)&--&--&12.2(1)\\
P9&9.2(2)&--&--&9.8(1)\\
P10&19.5(17)&--&--&13.1(1)\\
P11&9.7(7)&--&--&5.2(0)\\
P12&11.9(11)&--&--&14.9(2)\\
P13&14.3(40)&--&--&7.1(1)\\
P14&11.8(15)&--&--&7.7(0)\\
P15&17.5(3)&--&--&4.8(0)\\
P16&20.2(22)&--&--&18.6(3)\\
P17&30.2(5)&--&--&16.8(2)\\
P18&17.7(13)&--&--&7.4(1)\\
P19&20.2(5)&--&--&9.9(1)\\
P20&13.5(13)&--&--&29.3(1)\\
P21&7.3(9)&--&--&20.7(1)\\
P22&17.3(24)&--&--&20.3(1)\\
P23&8.5(1)&--&--&5.9(0)\\
P24&10.0(23)&--&--&15.7(2)\\
P25&7.1(6)&--&--&3.6(0)\\
P26&7.5(4)&--&--&6.0(0)\\
P27&11.4(8)&--&--&6.0(0)\\
P28&10.7(8)&--&--&14.8(1)\\
P29&14.0(4)&--&--&10.2(1)\\
P30&21.9(17)&--&--&8.7(1)\\
P31&19.0(14)&--&--&14.8(5)\\
P32&23.9(27)&--&--&8.0(0)\\
P33&9.9(3)&--&--&8.9(1)\\
P34&19.8(20)&--&--&25.8(3)\\
P35&15.2(76)&--&--&8.2(1)\\
P36&14.9(22)&--&--&15.9(1)\\
P37&33.1(7)&--&--&17.8(2)\\
P38&14.3(6)&--&--&16.3(0)\\
P39&12.2(18)&--&--&14.8(0)\\
P40&7.6(2)&--&--&9.0(1)\\
P41&16.8(5)&--&--&17.4(1)\\
P42&18.7(1)&--&--&35.4(1)\\
P43&12.2(15)&--&--&23.4(1)\\
P44&5.2(3)&--&--&9.6(0)\\
P45&7.2(7)&--&--&8.1(0)\\
P46&10.4(20)&--&--&22.4(0)\\
P47&7.8(2)&--&--&13.1(1)\\
P48&18.7(6)&--&--&17.9(0)\\
P49&11.1(7)&--&--&7.0(0)\\
P50&24.9(1)&--&--&38.0(2)\\
P51&22.0(1)&--&--&25.5(1)\\
P52&14.0(5)&--&--&13.8(0)\\
P53&6.0(0)&--&--&19.1(2)\\
P54&16.7(4)&--&--&10.1(0)\\
P55&11.9(5)&--&--&12.5(0)\\
P56&14.4(7)&--&--&13.1(0)\\
P57&13.4(17)&--&--&24.2(1)\\
P58&5.4(26)&--&--&18.0(0)\\
P59&9.6(29)&--&--&14.7(1)\\
P60&13.0(23)&--&--&16.1(0)\\
P61&10.3(18)&--&--&11.4(0)\\
P62&11.7(2)&--&--&19.0(1)\\
P63&18.4(35)&--&--&15.6(1)\\
P64&11.2(10)&--&--&18.5(0)\\
P65&11.0(17)&--&--&13.0(0)\\
P66&14.0(8)&--&--&24.9(1)\\
P67&2.0(1)&--&--&8.5(0)\\
P68&14.3(1)&--&--&40.4(0)\\
P69&14.4(1)&--&--&31.8(1)\\
P70&8.4(4)&--&--&16.1(0)\\
P71&10.3(3)&--&--&14.0(0)\\
P72&15.3(22)&--&--&16.8(1)\\
P73&13.6(4)&--&--&10.9(0)\\
P74&4.6(7)&--&--&11.4(0)\\
P75&11.2(2)&--&--&13.6(0)\\
P76&13.3(6)&--&--&10.1(0)\\
P77&20.6(6)&--&--&12.0(1)\\
P78&6.7(1)&--&--&12.9(0)\\
P79&14.6(5)&--&--&11.9(1)\\
\hline									
\textbf{VLA}&&\cite{2022Natur.606..873N}&&\\
L1\footnote{L-band (1-2 GHz)}&20&1.425&0.19 &--\\
L2&20&1.455&0.13 &--\\
L3&10&1.875&0.19 &--\\
S1\footnote{S-band (2-4 GHz)} &20&3.31&0.38 &--\\
S2&20&2.755&0.25 &--\\
S3&20&2.63&0.26 &--\\
S4&10&3.06&0.36 &--\\
S5&10&3&1.0 &--\\
Cl\footnote{C-band (5-7 GHz)}&10&4.685&0.39 &--\\
\hline									
\textbf{GBT}\footnote{C-band (4-8GHz)}&&\cite{2022Sci...375.1266F}&\\
1&--&4.82&--&--\\
2&--&4.92&--&--\\
3&--&5.51&--&--\\
\hline									
\textbf{Parkes}\footnote{the observing coverage [0.704-4.032] GHz} &&\cite{2022arXiv220308151D}&&\\
1&1.2&3.402&--&--\\
2&2.6&3.264&--&--\\
3&1.9&2.813&--&--\\
4&4.2&3.485&--&--\\
5&1.5&3.163&--&--\\
6&2.3&3.801&--&--\\
7&0.9&3.224&--&--\\
8&0.9&3.746&--&--\\
\hline									
\textbf{GBT}&&\cite{2022AAS...24022601T}&&\\
L1\footnote{L-band (1.1-1.9 GHz)}&11.7(3)&1.6&0.5&--\\
L2&4.2(1)&1.725&0.25&--\\
L3&5.4(4)&1.7&0.3&--\\
L4&2.3(2)&1.65&0.4&--\\
L5&3.4(4)&1.65&0.4&--\\
L6&9(1)&1.75&0.2&--\\
L7&23(3)&1.75&0.2&--\\
L8&3.7(5)&1.65&0.4&--\\
L9&6.0(9)&1.6&0.5&--\\
Cl.l\footnote{C-band (4-8 GHz)}&0.96(5)&4.312&0.377&--\\
C1.2&2.4(2)&4.33&0.34&--\\
C1.3&1.8(7)&4.482&0.257&--\\
C1.4&0.92(7)&4.33&0.34&--\\
C1.5&3.6(4)&4.33&0.34&--\\
C1.6&1.1(1)&5.75&0.5&--\\
C2.1&1.36(9)&4.875&0.75&--\\
C2.2&2.17(9)&5.434&2.633&--\\
C3.1&2.02(2)&4.792&1.317&--\\
C3.2&1.41(9)&4.792&1.317&--\\
C3.3&1.15(9)&4.887&1.126&--\\
C3.4&2.1(2)&4.7&1.5&--\\
C3.5&0.7(1)&6.2&0.75&--\\
C3.6&2.6(3)&4.513&1.125&--\\
C3.7&1.3(2)&4.7&0.75&--\\
C3.8&1.5(2)&5.45&0.75&--\\
\hline
\hline	
\end{longtable}


\bsp	
\label{lastpage}
\end{document}